\documentclass[runningheads]{llncs}
\usepackage[T1]{fontenc}
\usepackage{graphicx}
\usepackage{algorithm}
\usepackage{algorithmic}

\usepackage{booktabs}
\usepackage{multirow}
\usepackage[table,xcdraw]{xcolor}
\usepackage{tabularx}
\usepackage[utf8]{inputenc}
\usepackage{amsmath}

\begin{document}
\title{Inpainting Pathology in Lumbar Spine MRI with Latent Diffusion}
%
%
\author{Colin Hansen, Simas Glinskis, Ashwin Raju, Micha Kornreich, JinHyeong Park, Jayashri Pawar, Richard Herzog, Li Zhang, Benjamin Odry}
\authorrunning{Hansen et al.}
%
\institute{Covera Health, New York, NY, U.S.A. \\
\email{colin.hansen@coverahealth.com}}
\maketitle              
\begin{abstract}
Data driven models for automated diagnosis in radiology suffer from insufficient and imbalanced datasets due to low representation of pathology in a population and the cost of expert annotations. Datasets can be bolstered through data augmentation. However, even when utilizing a full suite of transformations during model training, typical data augmentations do not address variations in human anatomy. An alternative direction is to synthesize data using generative models, which can potentially craft datasets with specific attributes. While this holds promise, commonly used generative models such as Generative Adversarial Networks may inadvertently produce anatomically inaccurate features. On the other hand, diffusion models, which offer greater stability, tend to memorize training data, raising concerns about privacy and generative diversity. Alternatively, inpainting has the potential to augment data through directly inserting pathology in medical images. However, this approach introduces a new challenge: accurately merging the generated pathological features with the surrounding anatomical context. While inpainting is a well-established method for addressing simple lesions, its application to pathologies that involve complex structural changes remains relatively unexplored. We propose an efficient method for inpainting pathological features onto healthy anatomy in MRI through voxel-wise noise scheduling in a latent diffusion model. We evaluate the method’s ability to insert disc herniation and central canal stenosis in lumbar spine sagittal T2 MRI, and it achieves superior Fr\'echet Inception Distance compared to state-of-the-art methods. 

\end{abstract}
\section{Introduction}


Computer vision and natural language processing have progressed significantly in the last decade fueled by the numerous datasets comprised of millions of images, text segments, and the annotations that accompany them \cite{deng2009imagenet,schuhmann2021laion}. Medical image analysis has likewise benefited from these advances and has seen many breakthroughs in tasks such as segmentation and computer-aided diagnosis \cite{lundervold2019overview,shen2017deep}. Unlike many natural image tasks, acquiring MRI is both expensive and time consuming as is annotation which requires individuals with expertise in radiology and medicine. As a consequence, many MRI datasets merely consist of a few thousand samples and typically suffer from class imbalance and high inter-rater variability \cite{karimi2020deep}. \par

To compensate for these challenges, significant advancements have been made leveraging data augmentations to generalize models to features that would otherwise be missing from the training set \cite{shorten2019survey}. However, due to the complexity of individual human anatomy, it is difficult to develop transformations that create out-of-distribution examples that are also anatomically possible \cite{chlap2021review}. Additionally, augmentations which drastically alter an image will often remove or obstruct the image features relevant to a pathology. Recent advances in diffusion models (DMs) \cite{sohl2015deep,ho2020denoising} and latent diffusion models (LDMs) \cite{rombach2022high} have made image synthesis an attractive method for generating data through increased stability and model performance over generative adversarial networks (GANs). However, in the face of limited training data, DMs have been shown to be more likely than GANs to memorize and generate samples from the training set during inference \cite{carlini2023extracting,somepalli2023diffusion}. Regardless, a generative model trained on in-distribution data will generate in-distribution images.  In this work, we explore the use of inpainting to add pathological image features to normal samples to bolster classes with low representation. \par

On the other hand, inpainting presents an opportunity for building models capable of augmenting data while considering surrounding anatomy. However, typically inpainting uses regions of interest (ROI) or semantic segmentations to select features that should be replaced by a generative model. In medical imaging, using such ROIs can be challenging. For example, relevant features that define a pathology are often limited to small regions in the image and can involve multiple structures. If the ROI does not encompass these affected structures, the generative model will be incapable of synthesizing the pathology correctly. On the other hand, larger ROIs can lead to anatomically unrealistic images as anatomical priors become unavailable during generation. \par

In this paper, we propose a novel spatially weighted noise schedule for inpainting pathology using LDMs. To the best of our knowledge, this is the first time LDMs are utilized for inpainting pathological features into medical images with no pathology present while ensuring cohesive interaction with neighboring anatomical features. The experiments presented demonstrate significant improvements over existing baselines, highlighting the clinical relevance of our proposed method.



\begin{figure*}
    \centering
    \includegraphics[width=\textwidth]{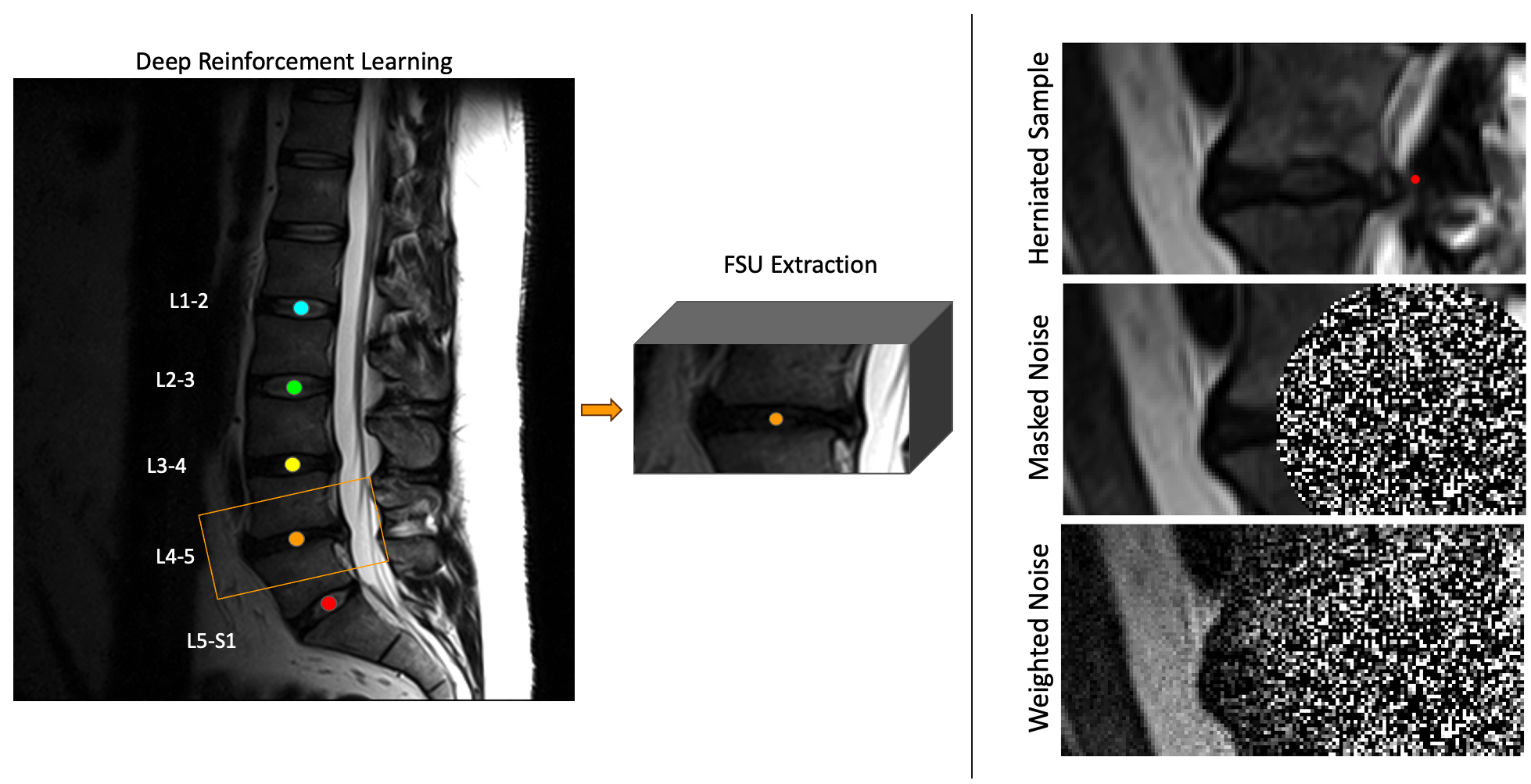}
    \caption{Left, a lumbar spine sagittal T2 weighted MRI is shown with L1-2 through L5-S1 disc centers localized by a deep reinforcement learning model. Associated FSUs are cropped to 8 $\times$ 8 $\times$ 5 cm. Right, a FSU with a disc herniation landmark (top) has noise added in a spherical ROI (middle) as well as Guassian weighted noise (bottom). 
    }
    \label{datafig}
\end{figure*}   

\section{Related Works}

\subsection{Image Synthesis}

Recent developments in image synthesis have opened up the possibility of creating imaging datasets generatively with high-resolution and involving complex scenes or objects. Diffusion models (DMs) have been shown to achieve state of the art results in image synthesis \cite{ho2020denoising,dhariwal2021diffusion} and video generation \cite{openai2024video} compared to autoregressive (AR) transformers which require billions of parameters \cite{razavi2019generating} and GANs which do not easily scale to modeling complex distributions. DMs are likelihood models and so do not suffer from mode-collapse or training instability \cite{sohl2015deep}. \par

\subsection{Diffusion Models}

Despite their advantages, DMs are still computationally demanding both during training and inference \cite{dhariwal2021diffusion}, and though many works have been developed to accelerate the training and inference processes while maintaining sample quality \cite{ho2020denoising,song2020denoising,liu2022pseudo}, when extending the DM framework to 3D data, the computational resources required may not be reasonably available. LDMs reduce computational requirements by learning the diffusion process in the latent space of an autencoder which downsamples the spatial dimensions of data without over expanding the feature space \cite{rombach2022high}. Recent works have shown that applying LDMs for MRI can produce high quality synthetic data conditioned on a variety of anatomically significant features \cite{pinaya2022brain}. Overcoming these computational and performance limitations are important steps towards fully synthesizing MRI datasets, but investigations have shown that with limited training data DMs can be twice as likely to reproduce training data rather than generate a unique sample when compared to their GAN counterparts \cite{carlini2023extracting}. \par

\subsection{Inpainting}
Inpainting, a branch of image synthesis, provides the ability to augment images only within an ROI to generate new samples in tandem with generative methods such as DMs \cite{lugmayr2022repaint}. Relying on a segmentation as the inpainting ROI is useful for replacing an object or background in an image \cite{liu2023grounding}. Segment anything model (SAM) \cite{kirillov2023segany} has been used to automate the annotation process for selecting an ROI on which to inpaint, and LLMs can guide and condition the generative process to inpaint based on a text input \cite{yu2023inpaint}. In the medical image domain, the time for an expert annotator to segment a structure or pathological region is quite expensive, and while SAM show great promise in scalability, its reliability in some medical tasks are still in need of investigation \cite{he2023accuracy}\cite{deng2023segment}. Annotating an anatomical landmark point is far less expensive than segmentation and more common. \par 
Inpainting has been explored for brain MRI \cite{manjon2020blind}\cite{liu2021symmetric} as well as lung CT \cite{astaraki2022prior} with the task of removing abnormal tissue to allow standard pipelines such as registration to operate under normal conditions. Lung nodule inpainting is also used to achieve higher accuracy for automatic screening \cite{yang2019class}. When inpainting nodules, lesions, or tumours onto healthy tissue or vice versa, it is often sufficient to modify the signal within a single anatomical structure. However, when it comes to inpainting pathologies which involve interactions between multiple structures, more complex methods need to be explored. \par

\section{Method}

\subsection{Latent Diffusion}

The LDM used in this work is adapted from Rombach et al. \cite{rombach2022high}, and the 3D implementation is adapted from Pinaya et al. \cite{pinaya2022brain}. The overall framework follows such that latent representation $z$ is obtained from passing input image $x$ to the encoder $\mathcal{E}$, and a U-Net $\epsilon_{\theta}$ is trained to model the diffusion process over $T$ timesteps within this latent space. At inference time decoder $D$ reconstructs the predicted image $\hat{x}$ from the estimated denoised latent representation $\hat{z}$. $\epsilon_{\theta}$ can be conditioned on a variety of information using encoder $\tau_{\theta}$ to attain an intermediate representation which is then mapped to the intermediate layers of the U-Net via a cross attention layer \cite{rombach2022high}. As long as the size of the feature space is constrained, the model benefits from the reduced spatial dimensions, but the framework depends on the performance of the decoder $D$ to attain high quality output. For encoder $\mathcal{E}$ and decoder $D$, we follow Rombach et al. and use a trained VQGAN autoencoder for all methods \cite{esser2021taming}. Our primary advancement is defining how and where the noise is added during the diffusion  process.

\subsection{Weighted Inpainting}

Rather than inpainting within the bounds of an ROI, our method sets a weighting scheme which modifies the current noise schedule timestep of each voxel given its distance from the anatomical landmark point. The weighting scheme is defined by replacing the landmark point with a  Gaussian sphere

\begin{equation}
    W=\frac{1}{\sigma\sqrt{2\pi}}e^{-||x-\mu||^{2}/2\sigma^{2}}
\end{equation}

with $\sigma=16$ mm as described by Kornreich et al. \cite{kornreich2022combining}. The timestep of each voxel $v$ can be written as

\begin{equation}
    t_{v}=[W_v*t]
\end{equation}
where $[\  ]$ denotes rounding to the nearest integer. If we let $\alpha_t=1-\beta_t$ and $\bar{\alpha}_t=\prod_{t}^{s=0}\alpha_s$ where $\beta$ is the variance of the Gaussian noise to be added to an image, then it follows that a noisy sample at a given voxel $x_{t_v}$ can be produced using the following distribution
\begin{equation}
    x_{t,v}\sim q(x_{t,v}|x_{0,v})=\mathcal{N}(x_{t,v}; \sqrt{\bar{\alpha}_{t_v}}x_{0,v}, (1-\bar{\alpha}_{t_v}))
\end{equation}
The model will learn to fully denoise the voxels near the landmark point and to partially denoise voxels based on their distance from the point as shown in Figure \ref{datafig}.

\subsection{Inpainting Pathology}

Our strategy for inpainting a pathology involves training only on data with the target pathology and an accompanying expert labeled landmark point. At inference time, we select data which an expert has annotated as not having this pathology, randomly choose a point from within a region where the pathology could possibly be present, and then apply the inpainting method using that point. Due to the trained model only having seen samples with pathology, it will generate pathological features which are anatomically possible given the surrounding context. At inference time we rely on pseudo numerical methods for diffusion models (PNDM) \cite{liu2022pseudo} to reduce the number of steps for generating images. 

\section{Experiments}

\subsection{Data}

Sagittal T2 sequences of MR lumbar spine were collected from 2,801 studies with an average voxel spacing of $0.625 \times 0.625 \times 4.3$ mm. 2,632 studies were acquired across three institutions with 34 institutions accounting for the remainder using hardware manufactured by GE, Siemens, Philips, Toshiba, and Hitachi. A magnetic field strength of 1.5 Tesla was used to acquire 2,033 studies followed by 3 Tesla which was used to acquire 504 studies with the remaining acquired using field strengths ranging from 0.6 to 1.2 Tesla. The studies were randomly split into train-validation-test sets in an approximately 70:10:20 ratio. Each functional spinal unit ((FSU), defined as a disc and two adjacent vertebral segments, were localized using a trained deep reinforcement learning model \cite{browning2021uncertainty} and then cropped to 8 $\times$ 8 $\times$ 5 cm as shown in Figure \ref{datafig}. An expert annotator labeled each disc herniation (DH) as either small or moderate to large, labeled each central canal stenosis (CCS) as moderate or severe, and provided a landmark annotation for both pathologies. All FSUs in the training and validation are used to train the autencoder which defines $\mathcal{E}$ and $D$ for the LDMs, and only the FSUs annotated with a herniation or stenosis are used to train the inpainting methods. In total there are 1,455 FSUs with DH landmarks and 3,699 with CCS landmarks (see Supplementary Table 1 for pathology distribution by FSU). The location of each FSU and the severity of the pathology are encoded as one hot vectors and used to condition the denoising process for all methods.

\subsection{Implementation Details}

Our network architecture implementation is adapted from Pinaya et al. \cite{pinaya2022brain}, and differs only in the autoencoder to account for the difference in image size and slice thickness. All methods use the same trained VQGAN autencoder which defines $\mathcal{E}$ and $D$. Due to the small image size after FSU extraction as well as a typical slice thickness of around 4.3 mm, the latent space of the autoencoder reduces the image dimensions from $1\times128\times64\times32$ to $3 \times 32 \times 16 \times 32$. The image slices are reduced by a magnitude of 4 while the number of slices is held constant, and only three features maps are used to encode the latent representation. The same architecture for the denoising U-Net $\epsilon_{\theta}$ is used for all methods as well. We use $T=1000$ for the diffusion and denoising process during training and set the number of inference steps to 50 using PNDM to reduce inference time. All models are trained with a batch size of 16 and a learning rate of $1e-4$ until convergence based on the validation loss, and models are trained for each pathology separately. Model training and inference were performed using PyTorch 2.0.1, MONAI 1.2.0, and an Nvidia 48 GB A6000 GPU.

\begin{table}[]
\centering
\caption{FID and MS-SSIM is reported for each method, pathology, and FSU. Red and yellow highlighting indicate low to moderate support in the target distribution.}
\resizebox{\textwidth}{!}{%
\begin{tabular}{llccccccc}
\hline
      &                                       & \multicolumn{1}{l}{} & \multicolumn{3}{c}{FID}                                                              & \multicolumn{3}{c}{MS-SSIM}            \\
FSU   & Condition                             & Support              & RePaint & Masked        & \multicolumn{1}{c|}{Weighted}                              & RePaint       & Masked & Weighted      \\ \hline
\rowcolor[HTML]{FF8F73} 
L1-2  & Mod./Large DH                         & 3                    & 0.41    & 0.38          & \multicolumn{1}{c|}{\cellcolor[HTML]{FF8F73}\textbf{0.32}} & 0.61          & 0.6    & \textbf{0.57} \\
\rowcolor[HTML]{FFF0B3} 
L2-3  & Mod./Large DH                         & 23                   & 0.63    & 0.58          & \multicolumn{1}{c|}{\cellcolor[HTML]{FFF0B3}\textbf{0.53}} & 0.5           & 0.48   & \textbf{0.46} \\
L3-4  & \cellcolor[HTML]{FFFFFF}Mod./Large DH & 88                   & 0.2     & 0.17          & \multicolumn{1}{c|}{\textbf{0.13}}                         & 0.52          & 0.5    & \textbf{0.47} \\
L4-5  & \cellcolor[HTML]{FFFFFF}Mod./Large DH & 129                  & 0.27    & 0.2           & \multicolumn{1}{c|}{\textbf{0.17}}                         & 0.5           & 0.48   & \textbf{0.45} \\
L5-S1 & \cellcolor[HTML]{FFFFFF}Mod./Large DH & 89                   & 0.26    & 0.2           & \multicolumn{1}{c|}{\textbf{0.15}}                         & 0.5           & 0.49   & \textbf{0.45} \\
\rowcolor[HTML]{FFF0B3} 
L1-2  & Severe CCS                            & 16                   & 0.47    & 0.36          & \multicolumn{1}{c|}{\cellcolor[HTML]{FFF0B3}\textbf{0.32}} & \textbf{0.66} & 0.7    & 0.67          \\
L2-3  & \cellcolor[HTML]{FFFFFF}Severe CCS    & 53                   & 0.47    & 0.45          & \multicolumn{1}{c|}{\textbf{0.37}}                         & 0.55          & 0.56   & \textbf{0.53} \\
L3-4  & \cellcolor[HTML]{FFFFFF}Severe CCS    & 160                  & 0.31    & 0.21          & \multicolumn{1}{c|}{\textbf{0.17}}                         & 0.55          & 0.56   & \textbf{0.54} \\
L4-5  & \cellcolor[HTML]{FFFFFF}Severe CCS    & 248                  & 0.81    & \textbf{0.54} & \multicolumn{1}{c|}{\textbf{0.54}}                         & 0.55          & 0.55   & \textbf{0.53} \\
\rowcolor[HTML]{FFF0B3} 
L5-S1 & Severe CCS                            & 21                   & 0.61    & \textbf{0.38} & \multicolumn{1}{c|}{\cellcolor[HTML]{FFF0B3}0.4}           & 0.47          & 0.43   & \textbf{0.41} \\ \hline
\end{tabular}%
}
\label{metrics}
\end{table}

\subsection{Baseline Methods}

We compare our method to two LDM baselines. The first is an LDM variant of RePaint \cite{lugmayr2022repaint}, where an LDM is trained to synthesize an entire image, and at inference time, each denoising step output $\hat{z}_{t-1}$ is modified such that the area outside of the ROI is replaced with the original latent with the noise from the previous timestep $z_{t-1}$. The second baseline trains the U-Net $\epsilon_{\theta}$ to denoise the image only within the defined ROI. Both baselines use the same ROI which is obtained by thresholding $W > 0.1$ to get a mask.

\subsection{Comparison}

All inpainting methods were trained to convergence and then applied to the testing set. The outputs of each method for test images without the pathology were compared to the distribution of test samples in which the pathology was found through the Fr\'echet Inception Distance (FID). This comparison is reported by FSU and conditioning. To further asses model performance, each method was also applied to test samples with the pathology present, and generation diversity  was assessed through MS-SSIM. Metrics were calculated by first cropping a $2.5 \times 2.5 \times 5$ cm patch surrounding the landmark point to limit the field of view to the inpainted region and then upsampling each cropped slice to an appropriate image size if required by the metric or pretrained network used to calculate the metric. \par
FID and MS-SSIM are reported in Table \ref{metrics} for the moderate to large DH and severe CCS pathologies. Metrics for small DH and moderate CCS are reported in Supplementary Table 2. The proposed weighted inpainting method outperforms the baselines in both FID and MS-SSIM for most FSUs and conditions. A visual comparison between methods is shown in Supplementary Figure 1. For qualitative clinical validation, we assess the performance of our method by providing an expert radiologist with 50 randomly selected inpainted FSUs from the test set for moderate to large DH as well as severe CCS. They are asked to rate each sample with a binary yes or no answer to if the image is anatomically realistic and if the image contains the target pathology. We report the percentage of images to which the answer was yes by FSU for DH and for CCS in Table \ref{clinical}. Across all 50 sampled FSUs, the radiologist found that  $86\%$ of samples inpainted with moderate to large herniations were realistic and $64\%$ contained the correct pathology. Likewise it was found that $82\%$ of samples inpainted with severe stenosis were realistic and $64\%$ contained the correct pathology.

\begin{table}[]
\centering
\caption{An expert radiologist rated randomly sampled FSUs inpainted with pathology using our method. We report the frequency in answering yes in determining if the image was realistic and if the desired pathology was present.}
\resizebox{\textwidth}{!}{%
\begin{tabular}{rcccccc}
\hline
\multicolumn{1}{l}{} & \multicolumn{3}{c}{Disc Herniation}                                                              & \multicolumn{3}{c}{Central Canal Stenosis}                                                                        \\ \hline
\rowcolor[HTML]{FFFFFF} 
FSU                  & Support & \%Realistic & \multicolumn{1}{c|}{\cellcolor[HTML]{FFFFFF}\%Mod./Large DH} & \cellcolor[HTML]{FFFFFF}Support & \cellcolor[HTML]{FFFFFF}\%Realistic & \cellcolor[HTML]{FFFFFF}\%Severe CCS \\ \hline
\rowcolor[HTML]{FFFFFF} 
L1-2                 & 7       & 85.71       & \multicolumn{1}{c|}{\cellcolor[HTML]{FFFFFF}42.86}                       & \cellcolor[HTML]{FFFFFF}10      & \cellcolor[HTML]{FFFFFF}80.00       & \cellcolor[HTML]{FFFFFF}40.00             \\
\rowcolor[HTML]{FFFFFF} 
L2-3                 & 14      & 85.71       & \multicolumn{1}{c|}{\cellcolor[HTML]{FFFFFF}57.14}                       & \cellcolor[HTML]{FFFFFF}6       & \cellcolor[HTML]{FFFFFF}83.33       & \cellcolor[HTML]{FFFFFF}33.33             \\
\rowcolor[HTML]{FFFFFF} 
L3-4                 & 8       & 87.50       & \multicolumn{1}{c|}{\cellcolor[HTML]{FFFFFF}75.00}                       & \cellcolor[HTML]{FFFFFF}7       & \cellcolor[HTML]{FFFFFF}85.71       & \cellcolor[HTML]{FFFFFF}71.43             \\
\rowcolor[HTML]{FFFFFF} 
L4-5                 & 11      & 90.91       & \multicolumn{1}{c|}{\cellcolor[HTML]{FFFFFF}81.82}                       & \cellcolor[HTML]{FFFFFF}14      & \cellcolor[HTML]{FFFFFF}85.71       & \cellcolor[HTML]{FFFFFF}78.57             \\
\rowcolor[HTML]{FFFFFF} 
L5-S1                & 10      & 80.00       & \multicolumn{1}{c|}{\cellcolor[HTML]{FFFFFF}60.00}                       & \cellcolor[HTML]{FFFFFF}13      & \cellcolor[HTML]{FFFFFF}76.92       & \cellcolor[HTML]{FFFFFF}76.92             \\ \hline
\end{tabular}%
}
\label{clinical}
\end{table}

\section{Discussion}
Inpainting can serve as a useful tool for data augmentation especially for datasets with imbalanced classes. In this paper, we demonstrate that the Gaussian weighted noise scheduling can improve inpainting performance. Our experiments show that the proposed method captures the complex interaction between the disc, vertebrae, and central canal for two pathologies, disc herniation and central canal stenosis. This is shown through the reduced FID and MS-SSIM suggesting the images inpainted with this technique are closer to the target distribution while being more diverse. \par
Notable limitations of this work involve hyperparameter optimization and data scarcity. The standard deviation for the Gaussian weighting was chosen ad hoc such that the noise sufficiently covered the structures that may be impacted by pathology. The optimal choice of this hyperparamater should be explored by pathology and perhaps could be varied depending on the intended size and impact of pathology. While inpainting holds promise for mitigating class imbalance issues in medical imaging, training on a relatively small dataset may hinder its performance and prevent it from reaching its full potential. All models generally performed worse in FSUs which had lower training support which is shown in both the quantitative metrics and the qualitative evaluation.

\section{Conclusion}

We have shown that weighted inpainting outperforms similar SOTA LDM methods. We have also shown that inpainting lumbar pathologies can be achieved with computationally efficient methods. Unique data augmentations such as these are necessary for medical imaging due to the complexity of clinical tasks and the variation in the human anatomy. 



\begin{credits}

\subsubsection{\discintname}
The authors have no competing interests to declare that are
relevant to the content of this article.
\end{credits}

\newpage
\bibliographystyle{splncs04}
\bibliography{main}
\newpage

\end{document}


\begin{table}[]
\centering
\caption{Distribution of pathologies by FSU. Notably, most pathology occurs in the L4-5 and then the L3-4 FSU.}
\begin{tabular}{rcccccc}
\hline
\multicolumn{1}{l}{} & \multicolumn{3}{c}{Disc Herniation}                                & \multicolumn{3}{c}{Central Canal Stenosis}                                                    \\ \hline
\rowcolor[HTML]{FFFFFF} 
FSU                  & Training & Validation & \multicolumn{1}{c|}{\cellcolor[HTML]{FFFFFF}Testing} & \cellcolor[HTML]{FFFFFF}Training & \cellcolor[HTML]{FFFFFF}Validation & \cellcolor[HTML]{FFFFFF}Testing \\ \hline
\rowcolor[HTML]{FFFFFF} 
L1-2                 & 46       & 11         & \multicolumn{1}{c|}{\cellcolor[HTML]{FFFFFF}6}       & \cellcolor[HTML]{FFFFFF}80       & \cellcolor[HTML]{FFFFFF}14         & \cellcolor[HTML]{FFFFFF}23      \\
\rowcolor[HTML]{FFFFFF} 
L2-3                 & 146      & 20         & \multicolumn{1}{c|}{\cellcolor[HTML]{FFFFFF}42}      & \cellcolor[HTML]{FFFFFF}369      & \cellcolor[HTML]{FFFFFF}49         & \cellcolor[HTML]{FFFFFF}105     \\
\rowcolor[HTML]{FFFFFF} 
L3-4                 & 247      & 31         & \multicolumn{1}{c|}{\cellcolor[HTML]{FFFFFF}89}      & \cellcolor[HTML]{FFFFFF}854      & \cellcolor[HTML]{FFFFFF}109        & \cellcolor[HTML]{FFFFFF}227     \\
\rowcolor[HTML]{FFFFFF} 
L4-5                 & 346      & 54         & \multicolumn{1}{c|}{\cellcolor[HTML]{FFFFFF}94}      & \cellcolor[HTML]{FFFFFF}1214     & \cellcolor[HTML]{FFFFFF}167        & \cellcolor[HTML]{FFFFFF}331     \\
\rowcolor[HTML]{FFFFFF} 
L5-S1                & 230      & 39         & \multicolumn{1}{c|}{\cellcolor[HTML]{FFFFFF}54}      & \cellcolor[HTML]{FFFFFF}107      & \cellcolor[HTML]{FFFFFF}23         & \cellcolor[HTML]{FFFFFF}27      \\ \hline
\end{tabular}
\end{table}

\begin{table}[]
\centering
\caption{FID and MS-SSIM is reported for each method, pathology, and FSU. Red and yellow highlighting indicate low to moderate support in the target distribution.}
\resizebox{\textwidth}{!}{%
\begin{tabular}{llccccccc}
\hline
      &                                      & \multicolumn{1}{l}{} & \multicolumn{3}{c}{FID}                                                              & \multicolumn{3}{c}{MS-SSIM}      \\
FSU   & Condition                            & Support              & RePaint & Masked        & \multicolumn{1}{c|}{Weighted}                              & RePaint & Masked & Weighted      \\ \hline
\rowcolor[HTML]{FF8F73} 
L1-2  & Small DH                             & 3                    & 0.74    & 0.67          & \multicolumn{1}{c|}{\cellcolor[HTML]{FF8F73}\textbf{0.62}} & 0.44    & 0.41   & \textbf{0.36} \\
\rowcolor[HTML]{FFF0B3} 
L2-3  & Small DH                             & 31                   & 0.81    & 0.74          & \multicolumn{1}{c|}{\cellcolor[HTML]{FFF0B3}\textbf{0.7}}  & 0.56    & 0.55   & \textbf{0.52} \\
L3-4  & \cellcolor[HTML]{FFFFFF}Small DH     & 93                   & 0.27    & 0.23          & \multicolumn{1}{c|}{\textbf{0.2}}                          & 0.52    & 0.5    & \textbf{0.47} \\
L4-5  & \cellcolor[HTML]{FFFFFF}Small DH     & 125                  & 0.42    & 0.34          & \multicolumn{1}{c|}{\textbf{0.3}}                          & 0.51    & 0.5    & \textbf{0.47} \\
L5-S1 & \cellcolor[HTML]{FFFFFF}Small DH     & 90                   & 0.32    & 0.27          & \multicolumn{1}{c|}{\textbf{0.22}}                         & 0.51    & 0.51   & \textbf{0.48} \\
\rowcolor[HTML]{FF8F73} 
L1-2  & Moderate CCS                         & 7                    & 0.41    & 0.4           & \multicolumn{1}{c|}{\cellcolor[HTML]{FF8F73}\textbf{0.33}} & 0.44    & 0.43   & \textbf{0.4}  \\
L2-3  & \cellcolor[HTML]{FFFFFF}Moderate CCS & 47                   & 0.63    & 0.68          & \multicolumn{1}{c|}{\textbf{0.58}}                         & 0.53    & 0.51   & \textbf{0.49} \\
L3-4  & \cellcolor[HTML]{FFFFFF}Moderate CCS & 59                   & 0.24    & 0.17          & \multicolumn{1}{c|}{\textbf{0.13}}                         & 0.49    & 0.47   & \textbf{0.45} \\
L4-5  & \cellcolor[HTML]{FFFFFF}Moderate CCS & 65                   & 0.53    & \textbf{0.29} & \multicolumn{1}{c|}{\textbf{0.29}}                         & 0.52    & 0.49   & \textbf{0.47} \\
\rowcolor[HTML]{FF8F73} 
L5-S1 & Moderate CCS                         & 5                    & 0.57    & \textbf{0.36} & \multicolumn{1}{c|}{\cellcolor[HTML]{FF8F73}0.42}          & 0.54    & 0.51   & \textbf{0.5}  \\ \hline
\end{tabular}%
}
\end{table}

\begin{figure*}[hbt!]
  \centering
  \includegraphics[width=1.0\textwidth]{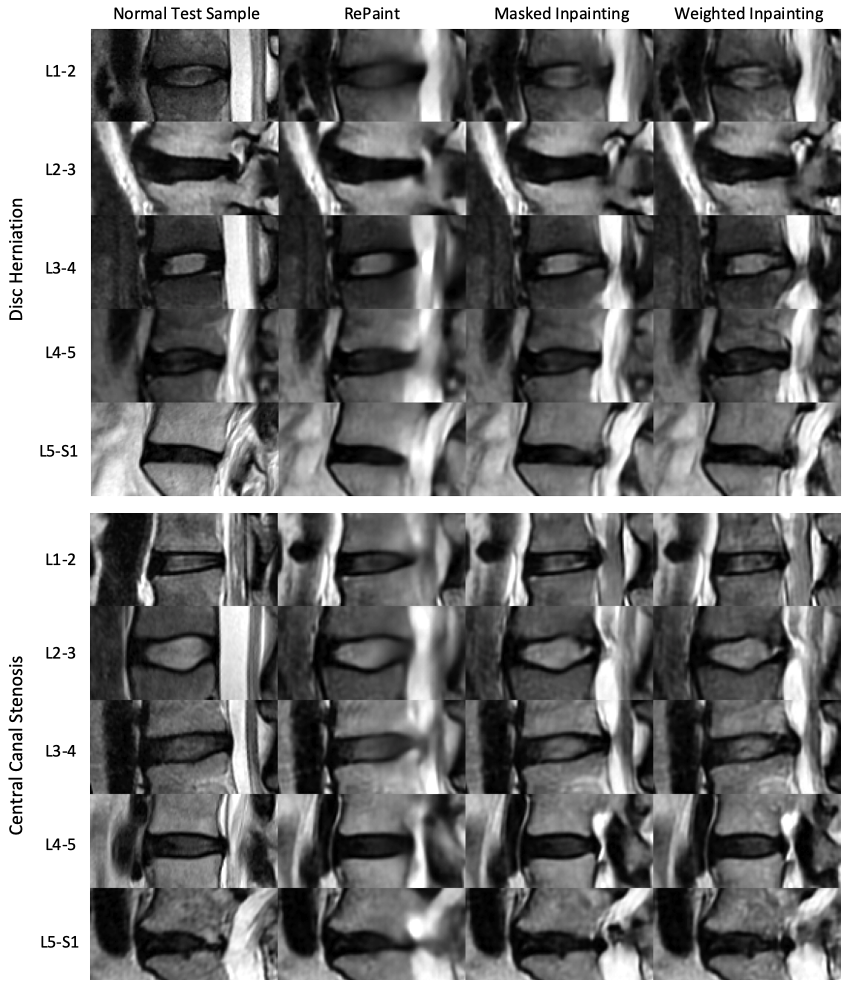}
  \caption{A qualitative assessment of inpainting performance between the three methods conditioned on moderate to large disc herniation (top) and severe central canal stenosis (bottom). The original normal motion segment from the test set (left) is shown along with the output pathology inpainting of each method: RePaint (center left), masked inpainting (center right), and weighted inpainting (right).}
\end{figure*}